\documentclass[onecolumn,preprintnumbers,amsmath,aps]{revtex4-1}
\usepackage{graphicx}
\usepackage{dcolumn}
\usepackage{bm}
\usepackage{subfigure}
\usepackage{color}
\usepackage{microtype}
\usepackage{amsmath}
\usepackage{mathtools}
\usepackage{amssymb}
\begin{document}
\def\eq#1{(\ref{#1})}
\def\fig#1{figure\hspace{1mm}\ref{#1}}
\def\tab#1{table\hspace{1mm}\ref{#1}}
\title{The hybrid cosmology in 
the scalar-tensor representation of $f(\mathcal{G},T)$ gravity}

\author{Adam Z. Kaczmarek}\email{a.kaczmarek@doktorant.ujd.edu.pl}
\author{Dominik Szcz{\c{e}}{\'s}niak}
\affiliation{Department of Theoretical Physics, Faculty of Science and Technology, Jan D{\l}ugosz University in Cz{\c{e}}stochowa, 13/15 Armii Krajowej Ave., 42200 Cz{\c{e}}stochowa, Poland}
\date{\today}
\date{\today} 
\begin{abstract}

In this work, the $f(\mathcal{G},T)$ theory of gravity is recast in terms of the $\phi$ and $\psi$ fields within the scalar-tensor formulation, where $\mathcal{G}$ is the Gauss-Bonnet term and $T$ denotes the trace of the energy-momentum tensor. The general aspects of the introduced reformulation are discussed and the reconstruction of the cosmological scenarios is presented, focusing on the so-called hybrid evolution. As a result, the scalar-tensor $f(\mathcal{G},T)$ theory is successfully reconstructed for the early and late time approximations with the corresponding potentials. The procedure of recovering the $f(\mathcal{G},T)$ theory in the original formulation is performed for the late time evolution and a specific quadratic potential. The scalar-tensor formulation introduced herein not only facilitates the description of various cosmic phases but also serves as a viable alternative portrayal of the $f(\mathcal{G},T)$ gravity which can be viewed as an extension of the well-established scalar Einstein-Gauss-Bonnet gravity.

\end{abstract}

\maketitle
\section{Introduction}

The dawn of a new era of cosmology was marked when the accelerated expansion of our Universe has been confirmed by different teams \cite{riess1998,perlmutter1999,eisenstein2005}, changing the long-standing former paradigms. Although important and insightful in unveiling new information about the Universe, this discovery caused problems for theoreticians mainly due the cosmological constant $\Lambda$ that has been brought back to life \cite{huterer2017}. The reintroduction of what Einstein once considered his {\it greatest mistake} serves now as a way to explain the accelerated expansion, playing role of the mysterious dark energy \cite{clifton2012}. Together with the cold dark matter (CDM), which is needed to explain lack of mass in the galactic halos and galaxy clusters, the $\Lambda$ constant is incorporated into the $\Lambda CDM$ model of cosmology. This picture, while properly describes accelerated expansion, is seriously challenged from not only theoretical standpoint but by the experiment as well \cite{Berti2015}. Noteworthy, the cosmological constant $\Lambda$ suffers from two main issues: the vacuum catastrophe and the coincidence problem \cite{copeland2006}. The mentioned questions led to the widespread belief that the $\Lambda CDM$ model is only an approximation of something much more fundamental \cite{abdalla2022}. This fact motivated many researchers to challenge the original formulation of the general relativity (GR), considering that Einstein's theory breaks down at large cosmological scales and some form of modification/extension to the gravitational action is needed \cite{clifton2012}. Interestingly, such attempts arised from the largerly forgotten ideas, often as old as the GR itself. Just to name a few, the first modifications of the GR introduced by Eddington, Weyl or Kaluza-Klein alike \cite{eddington1924,weyl1918,kaluza2018} or the follow-up concepts pioneered and developed over the several next decades \cite{brans1961,lovelock1971,Horndeski1974,stelle1977,damour1992}.

Hence, a vast landscape of different modified gravitational theories have been proposed over the recent years, including additional curvature invariants, coupling with the scalar and vector components or the non-trivial interactions between matter and curvature in the action principle \cite{clifton2012,lambiase2015}. One common approach is the $f(R)$ gravity, where the standard GR action is generalized to an arbitrary function of the Ricci scalar ($R \rightarrow f(R)$) \cite{buchdachl1970,nojiri2011,sebastiani2015f,malik2021,younesizadeh2021,gogoi2022}. Other curvature terms, such as the Gauss-Bonnet (GB) term, have been also included in the gravity's action as found within the $f(\mathcal{G})$ gravity \cite{nojiri2005,clifton2012,nojiri2021a}. In general, this kind of theories can be taxonomically grouped as the {\it higher-order theories}, since usually the extra couplings and curvature terms lead to the higher-order field equations \cite{nojrii2017}. It is important to add that the Gauss-Bonnet term arises naturally as a $\alpha'$-order correction to the effective action of the heterotic superstring theory \cite{zwiebach1985,metsaev1987,cano2022,fernandes2022}. Therefore, the modified gravity theories involving the GB term should be considered as much more sophisticated and complex than simple extensions of the GR created by adding another term into the considerations. In that manner, the $f(\mathcal{G})$ theory in fact can be reformulated as the scalar-Einstein-Gauss-Bonnet (scalar-EGB) gravity, linking it with the string-inspired dilatonic gravity \cite{nojiri2005,nojiri2005b}. As a result, a broad family of the $f(\mathcal{G})$ models have been successfully implemented in various aspects of cosmology, as they can {\it mimic} the dark sector (the dark energy and the dark matter) behavior and successfully describe inflation  \cite{capozziello2011,clifton2012,astashenok2015,zhong2018,kaczmarek2020,sahoo2020}. In general, obtaining cosmic evolution of the modified gravity is an uneasy task. To deal with such problem, the so-called reconstruction method have been implemented and commonly used in terms of the GR extensions \cite{nojiri2011,zubair2016,nojrii2017,odintsov2018,kaczmarek2020}. The main motivation behind such approach is the fact that by addition of the new degrees of freedom, the scale factor of interest is not determined uniquely by the matter content \cite{pinto2022}. For example, the de-Sitter evolution no longer requires inclusion of the dark energy, satisfying the main reasoning behind alternation of the Einstein's GR \cite{clifton2012,nojrii2017}. 

Recently, the higher-order theories have been broadly analyzed by taking into account (the non-minimal) curvature-matter couplings in the form of the $f(R,T)$ or the $f(R,L_m)$ gravity, where $L_m$ is the matter Lagrangian and $T$ denotes trace of the energy-momentum tensor $T_{\mu\nu}$ \cite{harko2011,myrzakulov2012,harko2014b,rudra2021,lobo2022,jaybhaye2022,pradhan2023}. The GB gravity has also been extended to the $f(\mathcal{G},T)$ theory by Sharif and Ikram \cite{sharif2016,sharif2017}. We note that consideration of the $T$-dependence can be motivated by the quantum effects, such as the conformal anomalies. Moreover, a nontrivial form of geometry-coupling have to be implemented if one wants to satisfy requirement of the conformal invariance for the natural laws \cite{scholz2018,lobo2022}. Hence, old works of Weyl \cite{weyl1918,weyl1952} should be treated as the main physical principles, guiding transition to the non-minimal couplings such as those in the $f(R,L_m)$ gravity \cite{lobo2022}. Besides that, the geometry-matter coupling manifests itself in a form of extra-force, influencing trajectories of the test particles. Moreover, the resulting general non-convariance of the energy momentum tensor ($\nabla_\mu T^{\mu\nu}\neq 0$) leads to the particle production by gravity \cite{harko2014b,pinto2022}. In terms of cosmological applications, models of that type have been extensively studied, leading to the important results as they can reproduce different cosmic models \cite{myrzakulov2012,moraes2017,harko2018e}. For example, in the context of the $f(\mathcal{G},T)$ gravity it was possible to reconstruct theory that is stable under linear perturbations \cite{sharif2017}. Cosmologically viable bouncing solutions have also been obtained, showing great flexibility of the $f(\mathcal{G},T)$ theory \cite{kaczmarek2020,shamir2021,yousaf2022}. Usefulness of the $f(\mathcal{G},T)$ gravity is also evident from other studies, such as those involving black holes or dealing with astrophysical applications \cite{bhatti2018,yousaf2020,maurya2020,zubair2021,sharif2022,sharif2022b,sharif2023}.

In this work, we reformulate the well-established $f(\mathcal{G},T)$ theory in terms of the equivalent biscalar Einstein-Gauss-Bonnet (EGB) gravity coupled with the trace of the energy-momentum tensor, being motivated by the connection between the scalar EGB and the $f(\mathcal{G})$ models. Such investigations are related to the fact that the Kaluza-Klein theories often lead to the multiple scalar fields coupled to the macroscopic distribution of energies, since compacton and dilaton are the remnants of the dimensional reduction of the spacetime \cite{damour1992}. Moreover, our work may be considered as another step in studying scalar-tensor representations of theories involving curvature-matter coupling, as scalar-tensor equivalent of the $f(R,T)$ gravity has been recently introduced and studied for the thin-shells, cosmology and particle production \cite{rosa2021,pinto2022,gonccalves2022}. Thus, after obtaining field equations and discussing the main characteristic of the new formulation, we use the reconstruction scheme to obtain different cosmic stages within the hybrid scale factor model given in \cite{akarsu2014}. Specifically, we focus on the early time and late time epochs that can be described by hybrid law in the corresponding limits as a power-law and de-Sitter like evolutions. Next, we delve into the analysis of the behavior of scalar field driving the cosmic epochs of interests. In the next step, our attention turns into recovering the $f(\mathcal{G},T)$ function from the scalar-tensor counterpart for the specific potentials $V(\phi,\psi)$. Our work ends with the summary of the obtained results, supported by comprehensive conclusions and future perspectives.

\section{Theoretical Model}

\subsection{Standard formulation}
The standard form of the general $f(\mathcal{G},T)$ gravity action given by the following relation, as postulated by Sharif and Ikram \cite{sharif2016}:
\begin{align}
    S=\frac{1}{2\chi}\int d^4x\sqrt{-g}\Big[ R+f\big(\mathcal{G},T\big)\Big] 
    + \int d^4x\sqrt{-g} \mathcal{L}_{m},
\label{eq1}
\end{align}
with $\chi=8\pi$ (for $G=c=1$) and $\mathcal{L}_m$ standing for the matter Lagrangian density, $g=\det(g_{\mu\nu})$. The functional $f(\mathcal{G},T)$ depends on the Gauss-Bonnet term $G=R^2-4 R_{\mu\nu}R^{\mu\nu}+R^{\mu\nu\alpha\beta}R_{\mu\nu\alpha\beta}$ and trace of the matter energy-momentum tensor $T=T^\mu_\mu=g^{\mu\nu}T_{\mu\nu}$. The energy-momentum tensor is defined as the variation of the $\mathcal{L}_m$ with respect to the metric $g_{\mu\nu}$ in the following manner:
\begin{align}
    T_{\mu\nu}=-\frac{2}{\sqrt{-g}}\frac{\delta (\sqrt{-g}\mathcal{L}_m)}{\delta g^{\mu\nu}}.
    \label{eq2}
\end{align}
Moreover, one can introduce auxiliary tensor $\Theta_{\mu\nu}$ as a form of variation of the $T_{\alpha\beta}$:
\begin{align}                           \Theta_{\mu\nu}=g^{\alpha\beta}\frac{\delta T_{\alpha\beta}}{\delta g_{\mu\nu}}.
\label{eq3}
\end{align}

By varying Eq.(\ref{eq1}), with respect to the metric tensor, we obtain the following field equations for the $f(\mathcal{G},T)$ gravity:
\begin{align}
\nonumber
    &G_{\mu\nu}+(T_{\mu\nu}+\Theta_{\mu\nu})f_T-\frac{1}{2}g_{\mu\nu}f +[2RR_{\mu\nu}-4R^\lambda_\mu R_{\lambda\nu}-4R_{\mu\alpha\nu\beta}R^{\alpha\beta}+2R^{\lambda\ \alpha\beta}_{\ \mu}R_{\nu\lambda\beta\alpha}]f_\mathcal{G}\\ 
    &+\big[2Rg_{\mu\nu}\Box-2R\nabla_\mu\nabla_\nu-4g_{\mu\nu}R^{\alpha\beta}\nabla_\alpha\nabla_\beta   -4R_{\mu\nu}\Box+4R^\lambda_\mu\nabla_\nu\nabla_\lambda+4R^\lambda_\nu\nabla_\mu\nabla_\lambda+4R_{\mu\alpha\nu\beta}\nabla^\alpha\nabla^\beta\big]f_\mathcal{G}=\chi T_{\mu\nu}.
\label{eq4}
\end{align}
Additionally, the trace of equation (\ref{eq4}) is given by:
\begin{align}
    R-(T+\Theta)f_T+2f+2\mathcal{G}f_\mathcal{G}-2R\Box f_\mathcal{G}+4R^{\mu\nu}\nabla_\mu\nabla_\nu f_\mathcal{G}+\chi T=0,
\label{eq5}
\end{align}
with $\Theta=\Theta^\mu_\mu$.

Next, the conservation equation for the $f(\mathcal{G},T)$ gravity can be obtained as a covariant divergence of the field equations (\ref{eq4}):
\begin{align}
    \nabla^\mu T_{\mu\nu}=\frac{1}{\chi - f_T}\Big[   (T_{\mu\nu}+\Theta_{\mu\nu})\nabla^\mu f_T+f_T \nabla^\mu \Theta_{\mu\nu}     -\frac{1}{2}\nabla_\nu f \Big].
\label{eq6}
\end{align}
Clearly the equation is nonvanishing. Moreover, we note that this is general behavior of theories with the nontivial couplings between geometry and matter \cite{harko2011,harko2014b,lobo2022}.
In what follows, the $f(\mathcal{G},T)$ gravity will be expressed in terms of tensor-scalar representation.

\subsection{Scalar-tensor formulation}

Now, we can introduce the scalar-tensor representation of the $f(\mathcal{G},T)$ theory as an extension of the scalar-tensor $f(\mathcal{G})$ gravity \cite{nojiri2005b}. In order to achieve this, we start with the two fields ($\alpha$ and $\beta$) in the following action principle:
\begin{align}
    S=\frac{1}{2\chi}\int d^4x\sqrt{-g}\Big[ R+f(\alpha,\beta)+(\mathcal{G}-\alpha)f_{\alpha}+(T-\beta)f_{\beta}\Big] + \int d^4x\sqrt{-g} \mathcal{L}_{m},
\label{eq7}
\end{align}
As in the scalar-tensor $f(R,T)$ gravity, the variation of the action Eq.(\ref{eq7}) with respect to the $\alpha$ and $\beta$ leads to the two equation of motion for these fields, that can be recast in the matrix form:
\begin{align}
 \textbf{M} \textbf{x}=   \begin{bmatrix}
f_{\alpha\alpha} & f_{\alpha\beta} \\
f_{\beta\alpha} & f_{\beta\beta} 
\end{bmatrix}  \begin{bmatrix}
\mathcal{G}-\alpha \\
T-\beta 
\end{bmatrix}=0,
\label{eq8}
\end{align}
where $f_{\alpha}=\frac{\partial f(\alpha,\beta)}{\partial{\alpha}}$ and so on. Moreover, we will assume that function $f(\alpha,\beta)$ satisfy the Schwartz theorem {\it i.e.} $f_{\alpha\beta}=f_{\beta\alpha}$. Then, the matrix form $\textbf{M} \textbf{x}=0$ has unique solution if $\det \textbf{M}\neq 0$, which translates to $f_{\alpha\alpha}f_{\beta\beta}\neq f^2_{\alpha\beta}$ \cite{pinto2022}. As a consequence, the unique solutions are $\alpha=\mathcal{G}$ and $\beta=T$, ensuring that action (\ref{eq7}) reduces to Eq.(\ref{eq1}) and provides direct equivalence between both of these formulations.

Finally, by defining scalar fields and potential $V(\phi,\psi)$ in the following manner:
\begin{align}
    \phi=\frac{\partial f}{\partial \mathcal{G}},\;\;\;\; \psi=\frac{\partial f}{\partial T},\;\;\;\; V(\phi,\psi)=\phi\alpha+\psi\beta -f(\alpha,\beta),
    \label{eq9}
\end{align}
the scalar-tensor representation of action (\ref{eq7}) (and as a consequence action (\ref{eq1})) reads:
\begin{align}
    S=\frac{1}{2\chi}\int d^4x\sqrt{-g}\Big[ R+\phi\mathcal{G}+\psi T - V(\phi,\psi)\Big] + \int d^4x\sqrt{-g} \mathcal{L}_{m}.
\label{eq10}
\end{align}
In this manner, we obtain the biscalar-EGB gravity with the non-trivial coupling between scalar field $\psi$ and matter via $\psi T$ term. In other words, action (\ref{eq10}) can be regarded as a specific scalar-EGB gravity extended by another degree of freedom \cite{nojiri2005,nojrii2017}. Now, variation with respect to the metric tensor $g_{\mu\nu}$ leads to the field equations:
\begin{align}
\nonumber
    G_{\mu\nu}+(T_{\mu\nu}+\Theta_{\mu\nu})\psi-\frac{1}{2}g_{\mu\nu}\big(\phi \mathcal{G}+ \psi T-V(\phi,\psi) \big)+[2RR_{\mu\nu}-4R^\lambda_\mu R_{\lambda\nu}-4R_{\mu\alpha\nu\beta}R^{\alpha\beta}+2R^{\lambda\ \alpha\beta}_{\ \mu}R_{\nu\lambda\beta\alpha}]\phi\\+\big[2Rg_{\mu\nu}\Box-2R\nabla_\mu\nabla_\nu-4g_{\mu\nu}R^{\alpha\beta}\nabla_\alpha\nabla_\beta-4R_{\mu\nu}\Box+4R^\lambda_\mu\nabla_\nu\nabla_\lambda+4R^\lambda_\nu\nabla_\mu\nabla_\lambda+4R_{\mu\alpha\nu\beta}\nabla^\alpha\nabla^\beta\big]\phi=\chi T_{\mu\nu},
\label{eq11}
\end{align}
while variation with respect to the $\phi$ and $\psi$ provides scalar equations:  
\begin{align}
    \mathcal{G}-\frac{\partial V}{\partial \phi}=0, \;\;\;\; T- \frac{\partial V}{\partial \psi}=0,
    \label{eq12}
\end{align}
for fields $\phi$ and $\psi$, respectively.
The four-divergence of the energy-momentum tensor for the scalar-tensor formulation reads:
\begin{align}
   \nabla^\mu T_{\mu\nu}=\frac{1}{\chi-\psi}\Big[(T_{\mu\nu}+\Theta_{\mu\nu})\nabla^\mu\psi-\frac{1}{2}\mathcal{G}\nabla_\nu \phi-\frac{1}{2}\nabla_\nu(\psi T-V)+\psi\nabla^\mu\Theta_{\mu\nu}\Big],
   \label{eq13}
\end{align}
and just like in the standard formulation of the $f(\mathcal{G},T)$ gravity, the energy-momentum is not generally conserved \cite{sharif2016,sharif2017}.

\subsection{General framework and modified Friedmann equations}

In order to study and reconstruct the hybrid-expansion that describes two different phases in the Universe history, we discuss basic assumptions and framework of the modified Friedmann-Lema{\^ i}tre-Robertson-Walker (FLRW) equations for the scalar-tensor $f(\mathcal{G},T)$ gravity. For the purposed of analyzing the isotropic and homogeneous Universe, we introduce the FLRW line element \cite{clifton2012,dodelson2020}:
\begin{align}
    ds^2=-dt^2+a^2(t)(dx^2+dy^2+dz^2),
    \label{eq14}
\end{align}
that is characterized by the scale factor $a(t)$.
Moreover, we assume a isotropic perfect fluid matter distribution for the Universe, characterized by the energy density $\rho$ and pressure $p$. The corresponding energy-momentum tensor is:
\begin{align}
    T_{\mu\nu}=(\rho+p)u_\mu u_\nu+p g_{\mu\nu},
\label{eq15}
\end{align}
where the four-velocity satisfies normalization $u^\mu u_\mu=-1$ and the matter Lagrangian is $L_m = p$.
Hence, tensor $\Theta_{\alpha\beta}$ is given by \cite{sharif2016}:
\begin{align}
    \Theta_{\mu\nu}=-2 T_{\mu\nu}+p g_{\mu\nu}.
\label{eq16}
\end{align}

Moreover, in order to preserve homogeneity and isotropy, all physical quantities: pressure, energy density and scalar fields $\phi$ and $\psi$, are assumed to be functions of the cosmic time $t$ alone. Hence, from the field equations Eq.(\ref{eq11}) for the line element Eq.(\ref{eq14}) one can obtain the Friedmann equations:
\begin{align}
    3H^2-\frac{1}{2}(3\rho-p)\psi+12H^3 \dot{\phi}+\frac{1}{2}\Big(\psi T-V(\phi,\psi)\Big)=\chi \rho,
\label{eq17}
\end{align}
and
\begin{align}
    -(2 \dot{H}+3H^2)-\frac{1}{2}\Big(\psi T-V(\phi,\psi)\Big)-8H(H^2+\dot{H})\dot{\phi}-4H^2 \Ddot{\phi}=\chi p,
\label{eq18}
\end{align}
where we have introduced Hubble rate $H=\dot{a}/a=\frac{da}{dt}/a$.
Moreover, the corresponding scalar equations (\ref{eq12}) are:
\begin{align}
24H^2(H^2+\dot{H})=\frac{\partial V}{\partial \phi},\;\;\;\;  -\rho+3p=\frac{\partial V}{\partial \psi}.
\label{eq19}
\end{align}
The conservation equation, on the other hand, takes the form:
\begin{align}
    \chi (\dot{\rho}+3H(\rho+p))=\psi\Big(\frac{1}{2}(3\Dot{\rho}-\dot{p})+3H(\rho+p)\Big)+\frac{\dot{\psi}}{2}(3\rho-p+V_\psi)+\frac{V_\phi}{2}\dot{\phi}-\frac{1}{2}\mathcal{G}\dot{\phi},
\label{eq20}
\end{align}
where shorthand notation for the potential derivatives $V_\phi=\frac{\partial V}{\partial \phi}$ and $V_\psi=\frac{\partial V}{\partial \psi}$ has been used.
The general nonvanishing four-divergence of the left-hand side of the field equations is a core motif in the theories involving non-trivial couplings between matter and geometry \cite{harko2014b,lobo2022}. Moreover, the purely geometrical part for the $V=V_1(\phi)+V_2(\psi)$ of the four-divergence vanishes, as expected in the $f(\mathcal{G})$ gravity. This is since in the limit $\psi \rightarrow 0$, the scalar-tensor $f(\mathcal{G},T)$ reduces to the usual scalar formulation of the $f(\mathcal{G})$ theory.
The following equations are not independent, in fact the system possess four linearly independent equations, similarly to the scalar tensor $f(R,T)$ gravity \cite{pinto2022,gonccalves2022}. Furthermore, by using relations for derivatives of potential $V$ one can eliminate both $V_\phi$ and $V_\psi$ leading to the four-divergence:
\begin{align}
\chi(\dot{\rho}+3H(\rho+p))=\psi\Big(\frac{1}{2}(3\dot{\rho}-\Dot{p})+3H(\rho+p)\Big)+\dot{\psi}(p+\rho).
\label{eq21}
\end{align}
Furthermore, we assume that perfect fluid satisfy the standard isotropic equation of state (EoS) \cite{dodelson2020}:
\begin{align}
    \frac{p(t)}{\rho(t)}=w,
\label{eq22}
\end{align}
with the parameter $w$ that varies depending on the matter content of the Universe. For instance, $w=-1$ translates to the dark matter (negative pressure), $w=1/3$ characterizes radiation while standard dust matter corresponds to the $w=0$.
In addition, we also assume that matter distribution is conserved in the sense of $\nabla^\mu T_{\mu\nu}=0$, which translates to:
\begin{align}
    \dot{\rho}+3H\rho(1+w)=0,
\label{eq23}
\end{align}
leading to the general form of the density and pressure:
\begin{align}
    \rho(t)=\rho_0 \Big(\frac{a}{a_0}\Big)^{-3(w+1)}, \;\;\; p(t)=w \rho(t),
\label{eq24}
\end{align}
with the integration constants $\rho_0$ and $a_0$.
Next, one can eliminate derivatives $\dot{\rho}$ and simplify the four-divergence to the following form:
\begin{align}
    \frac{1}{2} (w+1) \rho  \big(3 (w-1) H \psi+2
   \dot{\psi}\big)=0.
\label{eq25}
\end{align}
In the next section, the framework described above will be used to study cosmological scenarios from the hybrid-expansion law.

\section{Reconstruction of the cosmological scenarios}

\subsection{General solutions}

In the present section, we attempt to formulate and obtain general solutions for the auxiliary scale factor $a(t)$. First, it is important to observe that we are dealing with the eight unknowns ($a,\phi,\psi,\rho,p,V_\phi,V_\psi,w$). In this context, note that potential $V$ is a function of the two variables as the effective contribution comes from degrees of freedom associated with respectively $\phi$ and $\psi$, hence $V_\phi$ and $V_\psi$ are on the list. However, we already obtained expressions for the energy density and pressure given by Eq.(\ref{eq23}), allowing us to reduce the number of free parameters to six. Hence, we can simplify the whole system to the four independent equations {\it i.e.}  Eq.(\ref{eq17}), (\ref{eq19}), (\ref{eq20}). Once the scale factor $a(t)$ is determined, one can solve such system only with respect to the equation of state parameter $w$. Now we are ready to apply cosmological reconstruction procedure in the scalar-tensor framework of the $f(\mathcal{G},T)$ gravity.

By rewriting $H$ in terms of scale factors $a(t)$ and integration of the conservation equation leads to the general solution for $\psi(t)$:
\begin{align}
    \psi (t)= \psi_0 \frac{a(t)}{a_0}^{-\frac{3}{2} (w-1)}.
\label{eq26}
\end{align}
Since $\psi(t)$ is identical to the one obtained previously for $f(R,T)$, our reformulation of the $f(\mathcal{G},T)$ gravity is the correct one. Generally, four-divergence does not depend explicitly on the curvature invariants, being a {\it remnant} of the initial coupling with the trace of the energy-momentum tensor in the original formulation \cite{sharif2016,lobo2022}.

Moreover, one can invert relationship (\ref{eq26}) in order to get $a(\psi)$ dependence:
\begin{align}
   a(\psi)=a_0\big(\frac{\psi }{\psi_0}\big)^{\frac{2}{3-3 w}}.
\label{eq27}
\end{align}
Thus, scalar equation for $\psi$ is:
\begin{align}
    V_\psi=\rho_0 (3 w-1) \big(\big(\frac{\psi }{\psi _0}\big)^{\frac{2}{3-3 w}}\big)^{-3 w}
   \big(\frac{\psi }{\psi _0}\big)^{\frac{2}{w-1}}.
\label{eq28}
\end{align}
Clearly $V_\psi$ does not depend in any way on the scalar field $\phi$ coupled to the Gauss-Bonnet term. This characteristic implies that full auxiliary potential $V(\phi,\psi)$ is in fact separable in the variables $\psi$ and $\phi$. Thus, $V(\phi,\psi)$ will take form:
\begin{align}
V(\phi,\psi)=V_0+V_1(\phi)+V_2(\psi),
\label{eq29}
\end{align}
with the constant $V_0$. We notice that the nice split of potential given by  Eq.($\ref{eq29}$) is a common characteristic for the scalar-tensor formulations of the $f(\mathcal{G},T)$ and $f(R,T)$ gravities \cite{pinto2022,gonccalves2022}.
Thus, integration with respect to $\psi$ gives:
\begin{align}
  V_2(\psi)  =\frac{\psi  \rho_0 (w-1) (3 w-1) \big(\big(\frac{\psi }{\psi_0}\big)^{\frac{2}{3-3
   w}}\big)^{-3 w} \big(\frac{\psi }{\psi_0}\big)^{\frac{2}{w-1}}}{3 w+1}.
\label{eq30} 
\end{align}
Now, only the equations for $\phi$ and $V_\phi$ have to be solved:
\begin{align}\nonumber
  &\frac{12 a'(t)^3 \phi '(t)-a_0^3 \rho_0 \chi 
   \big(\frac{a(t)}{a_0}\big)^{-3 w}}{a(t)^3}+\frac{3 a'(t)}{a(t)}+a_0^3 \rho_0
   (w-2) \psi _0 a(t)^{-\frac{3}{2} (w+1)} \big(\frac{a(t)}{a_0}\big)^{-3 w}\\&+\frac{1}{2}
   \big(-\frac{\rho_0 \big(3 w^2-4 w+1\big) \psi _0 \big(a(t)^{-\frac{3}{2}
   (w-1)}\big)^{\frac{w+1}{w-1}} \big(\big(a(t)^{-\frac{3}{2} (w-1)}\big)^{\frac{2}{3-3
   w}}\big)^{-3 w}}{3 w+1}-V_1(\phi)-V_0\big)=0.
\label{eq31}
\end{align}
Furthermore taking the time derivative of Eq.(\ref{eq31}), by applying chain rule to the potential $\dot{V_1}(\phi)=\frac{\partial V_1}{\partial \phi}\dot{\phi}$ and invoking Eqs.(\ref{eq19}), leads to the equation:
\begin{align}\nonumber
 &-\frac{9}{4} \dot{a} a_0^3 \rho_0 w^2 \psi _0 a(t)^{-\frac{3}{2} (w-1)-4}
   \big(\frac{a(t)}{a_0}\big)^{-3 w}+\frac{3 \dot{a} a_0^3 \rho_0 \chi 
   \big(\frac{a(t)}{a_0}\big)^{-3 w}}{a(t)^4}+\frac{3 \dot{a} a_0^3 \rho_0 w \chi
    \big(\frac{a(t)}{a_0}\big)^{-3 w}}{a(t)^4}\\ \nonumber &+\frac{21}{4} \dot{a} a_0^3 \rho_0
   \psi _0 a(t)^{-\frac{3}{2} (w-1)-4} \big(\frac{a(t)}{a_0}\big)^{-3 w}+15 \dot{a} a_0^3
   \rho_0 w \psi _0 a(t)^{-\frac{3}{2} (w-1)-4} \big(\frac{a(t)}{a_0}\big)^{-3
   w}\\ \nonumber &+\frac{9 \dot{a} \rho_0 w^2 \psi _0 \big(a(t)^{-\frac{3}{2}
   (w-1)}\big)^{\frac{w+1}{w-1}} \big(\big(a(t)^{-\frac{3}{2} (w-1)}\big)^{\frac{2}{3-3
   w}}\big)^{-3 w}}{4 a(t)}-\frac{3 \dot{a} \rho_0 w \psi _0 \big(a(t)^{-\frac{3}{2}
   (w-1)}\big)^{\frac{w+1}{w-1}} \big(\big(a(t)^{-\frac{3}{2} (w-1)}\big)^{\frac{2}{3-3
   w}}\big)^{-3 w}}{a(t)}\\ &+\frac{3 \dot{a} \rho_0 \psi _0 \big(a(t)^{-\frac{3}{2}
   (w-1)}\big)^{\frac{w+1}{w-1}} \big(\big(a(t)^{-\frac{3}{2} (w-1)}\big)^{\frac{2}{3-3
   w}}\big)^{-3 w}}{4 a(t)}+\frac{24 \dot{a}^2 \dot{\phi } \ddot{a}}{a(t)^3}+\frac{12 \dot{a}^3
   \ddot{\phi }}{a(t)^3}-\frac{36 \dot{a}^4 \dot{\phi }}{a(t)^4}+\frac{3 \ddot{a}}{a(t)}-\frac{3
   \dot{a}^2}{a(t)^2}=0.
   \label{eq32}
\end{align}

\begin{figure}
    \centering
    \includegraphics[scale=1.3]{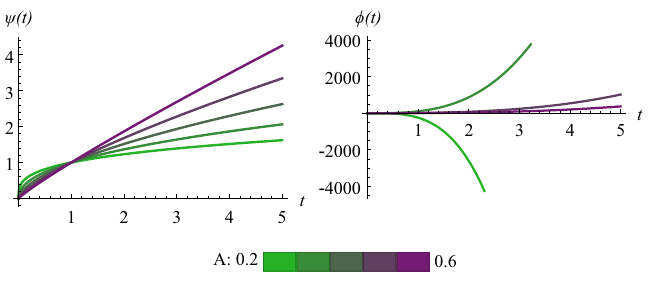}
    \caption{Time evolution of the scalar fields $\psi(t)$ and $\phi(t)$ in the early time limit of evolution (\ref{eq34}) for the different values of parameter $A\in \{0.2,0.3,0.4,0.5,0.6\}$. The constants used are: $\psi_1=t_0=a_0=\phi_0=1$ and $w=\phi_2=0$.}
    \label{fig:1}
\end{figure}

Then, from Eq.(\ref{eq32}), the general solution for the $\phi(t)$ will be the following:
\begin{align}\nonumber
&\phi (t)=\int _1^t\big(\frac{\phi _1
   a\big(t_2\big)^3}{a'\big(t_2\big)^3}+\frac{1}{a'\big(t_2\big)^3}\Big[\int _1^{t_2}\frac{1}{16}
   a\big(t_1\big)^{-\frac{3 w}{2}-4} \big(\frac{a\big(t_1\big)}{a_0}\big)^{-3 w}
   \big(\big(a\big(t_1\big)^{\frac{3}{2}-\frac{3 w}{2}}\big)^{-\frac{2}{3 (w-1)}}\big)^{-3
   w}\\ \nonumber &\big(-4 a_0^3 \rho_0 \chi  \big(\big(a\big(t_1\big)^{\frac{3}{2}-\frac{3
   w}{2}}\big)^{-\frac{2}{3 (w-1)}}\big)^{3 w} a'\big(t_1\big) a\big(t_1\big)^{3 w/2}-4
   a_0^3 w \rho_0 \chi  \big(\big(a\big(t_1\big)^{\frac{3}{2}-\frac{3
   w}{2}}\big)^{-\frac{2}{3 (w-1)}}\big)^{3 w} a'\big(t_1\big) a\big(t_1\big)^{3 w/2}\\ & \nonumber +16
   \big(\frac{a\big(t_1\big)}{a_0}\big)^{3 w}
   \big(\big(a\big(t_1\big)^{\frac{3}{2}-\frac{3 w}{2}}\big)^{-\frac{2}{3 (w-1)}}\big)^{3
   w} a'\big(t_1\big)^2 \phi '\big(t_1\big) a''\big(t_1\big) a\big(t_1\big)^{\frac{3
   w}{2}+1}+4 \big(\frac{a\big(t_1\big)}{a_0}\big)^{3 w}
   \big(\big(a\big(t_1\big)^{\frac{3}{2}-\frac{3 w}{2}}\big)^{-\frac{2}{3 (w-1)}}\big)^{3
   w}\\ \nonumber &a'\big(t_1\big)^2 a\big(t_1\big)^{\frac{3 w}{2}+2}-3 w^2 \rho_0
   \big(\frac{a\big(t_1\big)}{a_0}\big)^{3 w}
   \big(a\big(t_1\big)^{\frac{3}{2}-\frac{3 w}{2}}\big)^{\frac{w}{w-1}+\frac{1}{w-1}} \psi _0
   a'\big(t_1\big) a\big(t_1\big)^{\frac{3 w}{2}+3}+4 w \rho_0
   \big(\frac{a\big(t_1\big)}{a_0}\big)^{3 w}
   \big(a\big(t_1\big)^{\frac{3}{2}-\frac{3 w}{2}}\big)^{\frac{w}{w-1}+\frac{1}{w-1}}\\\nonumber &\psi _0
   a'\big(t_1\big) a\big(t_1\big)^{\frac{3 w}{2}+3}-\rho_0
   \big(\frac{a\big(t_1\big)}{a_0}\big)^{3 w}
   \big(a\big(t_1\big)^{\frac{3}{2}-\frac{3 w}{2}}\big)^{\frac{w}{w-1}+\frac{1}{w-1}} \psi _0
   a'\big(t_1\big) a\big(t_1\big)^{\frac{3 w}{2}+3}-4
   \big(\frac{a\big(t_1\big)}{a_0}\big)^{3 w}
   \big(\big(a\big(t_1\big)^{\frac{3}{2}-\frac{3 w}{2}}\big)^{-\frac{2}{3 (w-1)}}\big)^{3
   w}\\ \nonumber &a''\big(t_1\big) a\big(t_1\big)^{\frac{3 w}{2}+3}-7 a_0^3 \rho_0
   \big(\big(a\big(t_1\big)^{\frac{3}{2}-\frac{3 w}{2}}\big)^{-\frac{2}{3 (w-1)}}\big)^{3
   w} \psi _0 a'\big(t_1\big) a\big(t_1\big)^{3/2}+3 a_0^3 w^2 \rho_0
   \big(\big(a\big(t_1\big)^{\frac{3}{2}-\frac{3 w}{2}}\big)^{-\frac{2}{3 (w-1)}}\big)^{3
   w} \psi _0 a'\big(t_1\big) a\big(t_1\big)^{\frac{3}{2}}\\ &-20 a_0^3 w \rho_0
   \big(\big(a\big(t_1\big)^{\frac{3}{2}-\frac{3 w}{2}}\big)^{-\frac{2}{3 (w-1)}}\big)^{3
   w} \psi _0 a'\big(t_1\big) a\big(t_1\big)^{3/2}\big)dt_1
   a\big(t_2\big)^3\big)\Big]dt_2+\phi _2.
\label{eq33}
\end{align}

In principle, if $\phi(t)$ is invertible then by specifying $a(\phi)$ and using it in scalar equation (\ref{eq19}) one can get potential $V_1 (\phi)$ after integration.

\subsection{The hybrid expansion law}

For the purpose of studying various aspects of cosmology, we consider the so-called hybrid expansion model \cite{akarsu2014}:
\begin{align}
    a(t)=a_0 \Big(\frac{t}{t_0}\Big)^Ae^{B\big(\frac{t}{t_0}-1\big)},
\label{eq34}
\end{align}
characterized by the non-negative constants $A$ and $B$ as well as by the modern scale factor and the age of the Universe values of $a_0$ and $t_0$, respectively. The scale factor given by Eq.(\ref{eq34}) provides description of the power-law and the de Sitter evolutions in the limits of small and large times $t$ ({\it i.e.} in sense of the age of the Universe), as well as the transition between decelerated and accelerated stages of the cosmic history at $t=t_0(-A\pm \sqrt{A})$ for $0<A<1$ \cite{akarsu2014,koussour2022}.

From the simplified four-divergence of the $f(\mathcal{G},T)$ gravity, the time evolution of the scalar field $\psi(t)$ reads:
\begin{align}
\psi (t)= \psi_0 \exp{-\frac{3 (w-1) (A t_0 \log (t)+B t)}{2 t_0}}.
\label{eq35}
\end{align}
The relationship between $a(t)$ and $\psi(t)$ can be inverted to obtain $a(\psi)$ according to the previously derived relation (\ref{eq27}). Then, part of the potential associated with the trace of the energy-momentum $V_2(\psi)$ will be given by Eq.(\ref{eq26}).

Due to the complexity of the equations, we will assume that the Universe is filled with the dust-like matter, which allows us to take equation of state parameter $w=0$ and simplify the equations. Moreover, we will use the fact that the hybrid expansion law can be simplified due to its main characteristics - the description of both the de Sitter and the power-law evolutions at the different stages of history \cite{akarsu2014}. More specific, as discussed in \cite{akarsu2014}, for $B=0$ corresponding to the $t\rightarrow 0$ a power law phase occurs, while for late times $t \rightarrow \infty$ ($A=0$) the Universe will act approximately like the de-Sitter one. Hence, we can split our analysis for the early times and late times, taking into account different limits.

\begin{figure}
    \centering
    \includegraphics[scale=1.3]{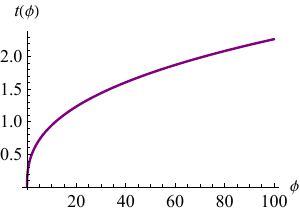}
    \caption{Numerical inverse of Eq.(\ref{eq37}) obtained for the early time limit of the scale factor (\ref{eq34}) given in Eq.(\ref{eq36})}
    \label{fig:2}
\end{figure}

\begin{figure}
    \centering
    \includegraphics[scale=0.8]{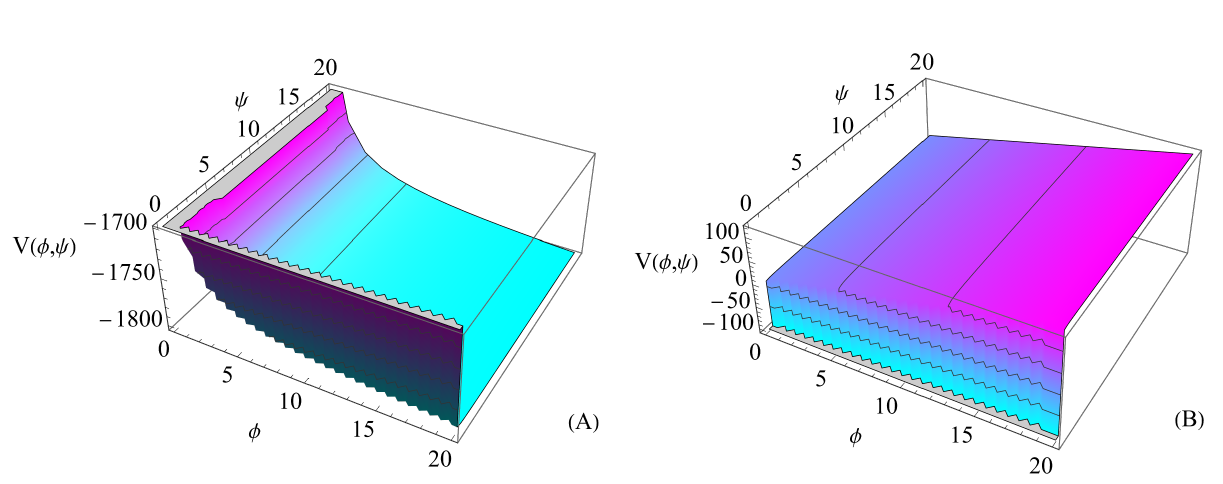}
    \caption{Potentials $V(\phi,\psi)$ in the early time (A) and late time (B) limits of the hybrid-law evolution. In plot (A) the $V_2(\psi)$ has been obtained analytically while $V_1(\phi)$ numerically. Plot (B) presents potential given by Eq.(\ref{eq43})} for $B=1/2$ and $t_0=1$.
    \label{fig:4}
\end{figure}

For the early time evolution $t\sim 0$, the scale factor $a(t)$ can be approximated as \cite{akarsu2014}:
\begin{align}
    a(t)\sim a_0\big(\frac{t}{t_0}\big)^A.
    \label{eq36}
\end{align}
Then, the scalar field $\phi(t)$, obtained from solving Eq.(\ref{eq31}), is found to be:
\begin{align}
    \phi (t)= -\frac{t^3}{12 A^3}-\frac{\big(7
   a_0^3+1\big) \rho_0 t^4 \psi _0
   \big(\frac{t}{t_0}\big)^{-2 A} \sqrt{a_0
   \big(\frac{t}{t_0}\big)^A}}{4 A^2 (3 A-8) (5
   A-2) a_0^2}-\frac{\rho_0 t^4 \chi 
   \big(\frac{t}{t_0}\big)^{-3 A}}{4 A^2 \big(12
   A^2-19 A+4\big)}+\frac{\phi_1
   t^{A+3}}{A+3}+\phi_2.
   \label{eq37}
\end{align}
Inversing potential in such a complicated form is an uneasy task that generally can be done numerically, if the conditions for invertibility are met {\it i.e.} $\phi(t)$ is bijective in nature. However, in the following scenario we are not able to go further by using pure analytic methods and inverse has to be performed numerically with the appropriate choice of constants and parameters. The numerically obtained inverse $t(\phi)$ of Eq.(\ref{eq37}) has been plotted in Fig.(\ref{fig:2}). Then, the inverse $t(\phi)$ can be used in the equation for the potential $V_1(\phi)$:
\begin{align}
    \frac{\partial V_1(\phi)}{\partial \phi}=24 \big(\frac{A^4}{t^4}-\frac{A^3}{t^4}\big),
    \label{eq38}
\end{align}
that can be numerically integrated in order to solve for $V_1(\phi)$. The full form of potential (\ref{eq29}) composed of Eq.(\ref{eq30}) and the solution of Eq.(\ref{eq39}) has been plotted in the left panel of Fig.(\ref{fig:4}) for the following set of parameters and constants: $(a_0=1;\rho_0=1,\psi_0=1,\phi_1=1,\phi_2=0,t_0=1,A=1/2,w=0)$.

\begin{figure}
    \centering
    \includegraphics[scale=1.3]{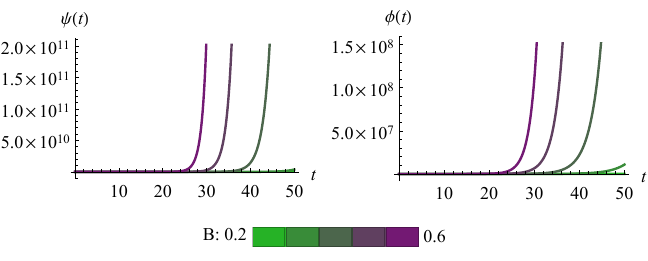}
    \caption{Visualization of the late time behaviour of the scalar fields $\psi(t)$ and $\phi(t)$ for the different values of $B\in\{0.2,0.3,0.4,0.5,0.6\}$, where $\psi_1=t_0=a_0=\phi_0=1$ and $w=\phi_2=0$. }
    \label{fig:3}
\end{figure}

For the different situation of late times, the exponential term dominates the scale-factor, in the de-Sitter-like evolution \cite{akarsu2014}:
\begin{align}
    a(t)\rightarrow e^{B(\frac{t}{t_0}-1)}, \;\;\; \text{for}\;\;\; t\rightarrow \infty.
    \label{eq39}
\end{align}
In the scenario characterized by the scale-factor Eq.(\ref{eq39}) one gets the following solution of the Eq.(\ref{eq31}):
\begin{align}
  \phi (t)=\frac{\rho_0 t_0^4 e^{B
   \big(2-\frac{3 t}{t_0}\big)} \big(16 a_0 C_1
   e^{\frac{B t}{t_0}} \sqrt{a_0 e^{B
   \big(\frac{t}{t_0}-1\big)}}+5 e^B \chi \big)}{240
   B^4}+\frac{C_2 t_0 e^{\frac{B
   t}{t_0}}}{B}+\phi_0,
   \label{eq40}
\end{align}
with the integration constants $\phi_0,C_1,C_2$. Behavior of the fields $\psi$ from Eq.(\ref{eq35}) ($A=0$) and $\phi$ from Eq.(\ref{eq40}) in the late time limit is presented in \ref{fig:3}. Now, only the equation for $V_1(\phi)$ part of the potential has to be solved. Since at the late time limit the scalar equation (\ref{eq12}) for $\phi$ is:
\begin{align}
    V_\phi=\frac{24 B^2}{t_0^4},
    \label{eq41}
\end{align}
being constant independent on $t$, the $V_\phi$ does not depend explicitly on $\phi$.  Thus, direct integration of Eq.(\ref{eq41}) leads to the solution:
\begin{align}
    V_1(\phi)=\frac{24 B^2}{t_0^4}\phi,
     \label{eq42}
\end{align}
where integration constant is absorbed by $V_0$. Hence, for the de-Sitter-like evolution, the biscalar potential is linear in $\phi$. Interestingly, this kind of dependence is similar to the result obtained in the scalar-tensor $f(R,T)$ situation \cite{gonccalves2022}. Hence, in the late time limit, the full potential takes form:
\begin{align}
    V(\phi,\psi)=V_0+\frac{24B^2}{t_0^4}\phi-\frac{a_0^3 \psi_0^2 \rho_0}{\psi },
    \label{eq43}
\end{align}
where Eq.(\ref{eq30}) for $w=0$ has been taken into account. The evolution of potential $V(\phi,\psi)$ from (\ref{eq43}) is depicted in the (B) panel of Fig(\ref{fig:4}).

\begin{figure}
    \centering
    \includegraphics[scale=0.8]{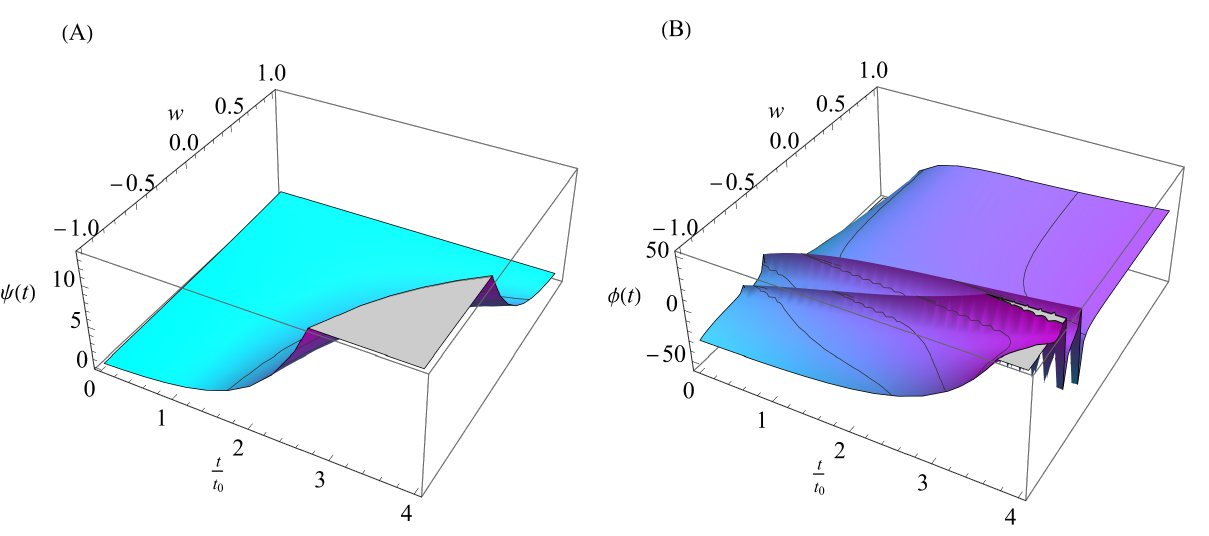}
    \caption{Late time evolution of the scalar fields $\psi(t)$ and $\phi(t)$ for different values of the equation of state parameter $w\in \langle-1,1\rangle$.}
    \label{fig:5}
\end{figure}

As a closing remark for this section, we turn our attention to the general scenario where $w$ can be arbitrary, and try to answer the following question: how the scalar field will behave now? For instance, the scalar field $\psi(t)$ that drives the de-Sitter-like scale factor is: 
\begin{align}
 \psi(t)=\psi _0 \big(e^{\frac{1}{2} (\frac{t}{t_0}-1)}\big)^{-\frac{3}{2} (w-1)}.
 \label{eq44}
\end{align}
while field $\phi(t)$ from Eq.(\ref{eq31}) reads:
\begin{align}\nonumber
    \phi (t)&= \frac{t_0}{12 B^4}
   \big(-\frac{\rho_0 t_0^3 \big(3 w^2-4
   w+1\big) \psi _0 \big(\big(a_0 e^{B
   \big(\frac{t}{t_0}-1\big)}\big)^{-\frac{3}{2}
   (w-1)}\big)^{\frac{w+1}{w-1}} \big(\big(\big(a_0
   e^{B \big(\frac{t}{t_0}-1\big)}\big)^{-\frac{3}{2}
   (w-1)}\big)^{\frac{2}{3-3 w}}\big)^{-3 w}}{(3 w+1) (9
   w+5)} \\ \nonumber
   &+\frac{a_0 \rho_0 t_0^3 (w-7) \psi
   _0 e^{B \big(2-\frac{2 t}{t_0}\big)} \big(e^{B
   \big(\frac{t}{t_0}-1\big)}\big)^{-3 w}
   \big(a_0 e^{B
   \big(\frac{t}{t_0}-1\big)}\big)^{\frac{1}{2} (1-3
   w)}}{9 w+5}+12 B^3 c_1 e^{\frac{B
   t}{t_0}}\\ &-\frac{\rho_0 t_0^3 \chi  e^{B
   \big(3-\frac{3 t}{t_0}\big)} \big(e^{B
   \big(\frac{t}{t_0}-1\big)}\big)^{-3 w}}{3
   w+4}\Big)+\phi_2.
   \label{eq45}
\end{align}
Although behavior of the field $\psi$ is smooth across all the values of $w$, it is not generally a case for the field $\phi$ since discontinuities occur when either of the denominators in Eq.(\ref{eq45}) is equal to $0$. This situation may correspond to the specific values of $w$, as can be seen in Fig.(\ref{fig:5}). Fortunately, as we are dealing with the modification of the GR, the evolution of the scale factor is not confined to the specific value of equation of state parameter $w$. For example, the de-Sitter evolution can be described for the dust ($w=0$) or radiation ($w=1/3$), contrary to the GR, where cosmological constant $\Lambda$ characterized by $w=-1$ is responsible for the accelerated phase \cite{dodelson2020}.

\section{Recovering specific $f(\mathcal{G},T)$ model}

From the previous discussion given in the section I it is evident that one can relate the obtained potential $V(\phi,\psi)$ to the corresponding $f(\mathcal{G},T)$ functional. In order to achieve this goal, we invoke definitions from Eqs.(\ref{eq9}), resulting in the partial differential equation for the function $f(\mathcal{G},T)$:
\begin{align}
    f(\mathcal{G},T)=-V(f_{\mathcal{G}},f_T)+f_{\mathcal{G}}\mathcal{G}+f_T T.
    \label{eq46}
\end{align}
Now, one can take the partial derivatives of Eq.(\ref{eq46}) with respect to the GB term and trace of the energy-momentum tensor, leading to the system of two partial differential equations:
\begin{align}
f_{\mathcal{G}\mathcal{G}}(\mathcal{G}-V_\phi)+f_{\mathcal{G}T}(T-V_\psi)=0,\;\;\;\; f_{T\mathcal{G}}(\mathcal{G}-V_\phi)+f_{TT}(T-V_\psi)=0,
\label{eq47}
\end{align}
or equivalently:
\begin{align}
    \begin{bmatrix}
f_{\mathcal{G}\mathcal{G}} & f_{\mathcal{G}T} \\
f_{T\mathcal{G}} & f_{TT} 
\end{bmatrix}  \begin{bmatrix}
\mathcal{G}-V_\phi \\
T-V_\psi
\end{bmatrix}=0.
\label{eq48}
\end{align}
This equation can be satisfied for a auxiliary function the $f(\mathcal{G},T)$ since the scalar equations for $\phi$ and $\psi$ lead to the following association: $V_\phi=\mathcal{G}$ and $V_\psi=T$, and cancellation of the left hand side of Eqs.(\ref{eq47}). Then, by using definitions $\phi=f_{\mathcal{G}}$ and $\psi=f_T$ and integrating independently, one can obtain function $f(\mathcal{G},T)$. Afterwards, upon inserting the solution into the Eq.(\ref{eq46}) one will arrive at the precise geometrical form of the original formulation, as some constraints on the integration constants may be required \cite{sharif2016,gonccalves2022}.

To demonstrate the above procedure, we will consider late time approximation of the hybrid scale factor from Eq.(\ref{eq39}). Despite the fact that the analysis for the early times has been also conducted in previous section, $\phi$-part of potential $V(\phi,\psi)$ have been obtained only numerically, because analytic expressions became too complex. Hence, in order to demonstrate the procedure of recovering geometric $f(\mathcal{G},T)$ gravity, we will focus our attention on the late time limit of hybrid model \cite{akarsu2014}. Then. one can rewrite the equation for the $\psi$-part of the potential:
\begin{align}
    V_\psi=T=\rho_0 (3 w-1) \big(\big(\frac{f_T}{\psi _0}\big)^{\frac{2}{3-3 w}}\big)^{-3 w}
   \big(\frac{f_T}{\psi _0}\big)^{\frac{2}{w-1}}.
\label{eq49}
\end{align}
Note that since Eq.(\ref{eq49}) does not depend explicitly on the GB term $\mathcal{G}$, the final form of the $f(\mathcal{G,T})$ is in fact separable in variables $\mathcal{G}$ and $T$, thus $f(\mathcal{G},T)=f_0 +f_1(\mathcal{G})+f_2(T)$. Note that this fact directly corresponds to the similar behavior of potential $V(\phi,\psi)$ in the scalar-tensor form. Inversion of Eq.(\ref{eq49}) for $f_T$ (i.e. $f_{T}$) leads to the:
\begin{align}
    f_T=
   \big(\frac{\rho_0 (3 w-1)
   \psi_0^{-\frac{2
   (w+1)}{w-1}}}{T}\big)^{\frac{1-w}{2
   w+2}}.
   \label{eq50}
\end{align}
For $w=\{-1,1/3\}$, Eq.(\ref{eq50}) has no explicit dependence on the $f_T$, hence the functional $f(\mathcal{G},T)$ depends on $T$ in arbitrary manner \cite{gonccalves2022}. In the scenario where $w=\neq\{-1,1/3\}$, the integral of Eq.(\ref{eq50}) with respect to $T$ gives:
\begin{align}
    f_2(T)=\frac{2 T (w+1) \big(\frac{\text{$\rho
   $0} (3 w-1) \psi_0^{-\frac{2
   (w+1)}{w-1}}}{T}\big)^{\frac{1-w}{2
   w+2}}}{3 w+1},
   \label{eq51}
\end{align}
and for $w=0$ it reduces to the $f_2(T)=\pm 2\psi_0 (-\rho_0 T)^{1/2}$, where $\pm$ arises due to the square root. We note that care should be taken, since solution (\ref{eq50}) is undefined for the equation of state parameter $w=1$. Note, that minus sign in front of the $\rho_0$ cancels with the $T<0$ {\it i.e.} for $\big(\rho <0\land w>\frac{1}{3}\big)$ or $\big(\rho >0\land w<\frac{1}{3}\big)$. Remarkably, the same functional behavior has been obtained in the scalar-tensor $f(R,T)$ gravity for the de-Sitter evolution with the equation of state parameter $w=0$ \cite{gonccalves2022}. This fact also verifies our analysis, as both the $f(R,T)$ and the $f(\mathcal{G},T)$ theories should result in the $f(T)$ models for $f(R,T)\rightarrow R+f(T)$ and $f(\mathcal{G},T)\rightarrow f(T)$, respectively \cite{harko2014b,sharif2016}.

Now, let us turn attention to the dependence on the GB term. For the late time limit and with respect to the considered hybrid evolution law, the analytical form of $V_\phi$ from (\ref{eq39}) had no explicit dependence on the scalar field $\phi$. Similarly to the situation for $f_T$ with the choice $w=\{-1,1/3\}$, dependence on the GB term $\mathcal{G}$ is unconstrained. Then, for the general situation, the $f(\mathcal{G},T)$ can be arbitrary in the GB term dependence, or if the solution in $T$ given by Eq.(\ref{eq48}) is valid, it will be arbitrary in $f_1(\mathcal{G})$.
Hence, functional form that describes late time scale factor from Eq.(\ref{eq39}) for $w=0$ will be:
\begin{align}
f(\mathcal{G},T)=f_0+f_1(\mathcal{G})\pm 2\psi_0 (-\rho_0 T)^{1/2}.
    \label{eq52}
\end{align}
In order to fix the constant $f_0$, we combine Eq.(\ref{eq52}) with the potential given by Eq.(\ref{eq43}) in the Eq.(\ref{eq46}), leading to the $f_0=-V_0$.

In order to better grasp the transition from the scalar-tensor to the geometrical (standard) formulation of the $f(\mathcal{G},T)$ gravity, we will consider one specific toy-model for potential $V(\phi,\psi)$, that describes some scale factor $a(t)$, in the form of simple additive quadratic potential \cite{pinto2022}:
\begin{align}
    V(\phi,\psi)=\alpha \phi+\beta \phi^2 -\gamma \psi^2.
    \label{eq53}
\end{align}
The above has been used for instance in the scalar-tensor $f(R,T)$ description of matter-production \cite{pinto2022}. Then, from Eqs. (\ref{eq12}) and with $\phi=f_\mathcal{G}$ and $\psi=f_T$ one gets:
\begin{align}
    f_\mathcal{G}=\frac{1}{2\beta}(\mathcal{G}-\alpha),\;\;\; \text{and} \;\;\; f_T=-\frac{T}{2\gamma}.
    \label{eq54}
\end{align}
Note that again, functional $f(\mathcal{G},T)$ takes the separated form of $f(\mathcal{G},T)=f_0+f_1(\mathcal{G})+f_2(T)$, as the potential is separated in $\phi$ and $\psi$. Expressions (\ref{eq54}) can be integrated with respect to the GB term and trace of the energy-momentum tensor respectively, yielding:
\begin{align}
    f_1(\mathcal{G})=\frac{1}{2\beta}(\frac{1}{2}\mathcal{G}^2-\alpha\mathcal{G}),\;\;\;\; f_2(T)=-\frac{T^2}{4\gamma}.
    \label{eq55}
\end{align}
Now only $f_0$ is left to be determined, what can be done by using potential (\ref{eq53}) together with the Eqs. (\ref{eq54}) and (\ref{eq55}) in the Eq.(\ref{eq46}). After straightforward manipulation, one gets $f_0=\alpha^2/(4\beta)$, completing the transition from the scalar-tensor to the original formulation of the $f(\mathcal{G},T)$ gravity for the potential defined by Eq.(\ref{eq53}).

\section{Summary and conclusions}

In the summary, the purpose of our work was two-fold, as we have focused on the construction of scalar-tensor formulation of the $f(\mathcal{G},T)$ gravity. By following this process, we have also obtained different cosmological scenarios from the hybrid scale factor given by Eq.(\ref{eq34}). As a resluts, we have observed that, in the framework of the scalar-tensor formulation, the resulting equations of motion are at most the second order, contrary to the higher order expressions of the original formulation presented in \cite{sharif2016}. In what follows, in principle it is easier to differentiate specific contributions for the matter-geometry couplings as every extra term to the Einstein field equations is either coupled with the $\phi$ or $\psi$. Then, the resulting framework has been applied into the Friedmann-Lema{\^ i}tre-Robertson-Walker (FLRW) metric in the Cartesian coordinates, where main characteristics of the theory has been investigated. This included the general form of the scalar fields $\psi(t)$ and $\phi(t)$ and procedure for obtaining evolution that satisfies the energy-momentum conservation equation, which is not necessary a case in the theories of that kind \cite{harko2014b,sharif2016,lobo2022,rosa2021}. Next, the reconstruction procedure introduced here has been applied for the scale-factor that may describe different cosmic phases, namely for the hybrid-evolution law \cite{akarsu2014}: $a(t)=a_0 \Big(\frac{t}{t_0}\Big)^Ae^{B\big(\frac{t}{t_0}-1\big)}$ in the two limits corresponding to the different cosmic eras. In the early time limit, this corresponds to the specific power-law evolution $a(t)\rightarrow  a_0\big(\frac{t}{t_0}\big)^A$, where we have obtained potential $V(\phi,\psi)$ using the combined numerical and analytical procedure. Then, our analysis involved reconstruction of the scalar-tensor form of the $f(\mathcal{G},T)$ gravity in the late-time limit of the hybrid evolution. This investigation led to the derivation of a fully analytical potential $V(\phi,\psi)$ for late times.

However, a few points have to be taken into account. Firstly, main difficulties may arise when $\phi$ and $\psi$, as a functions of the metric and associated coordinates, are not invertible. While we were able to obtain their inverse forms numerically, it may still not be the case in some circumstances. Another remark is that ambiguities might arise during the reconstruction procedure, since one can encounter discontinuities associated with the specific parameters values. However, akin to similar issues observed in the case of the equation of state parameter $w$ in the scalar-tensor $f(R,T)$ theory, these can be avoided altogether \cite{gonccalves2022}. To be more precise, for the scalar-tensor $f(\mathcal{G},T)$ one can always assume a constant equation of state through all cosmic times $t$, since all changes in the evolution of the scale factor can be associated with the terms and degrees of freedom from the modified gravity of interest. In other words, the contribution from the scalar-tensor $f(\mathcal{G},T)$ theory can be effectively treated as dark energy or dark matter \cite{sharif2016}. On the other hand, the discontinuities in $\psi$, $\phi$ and $V(\phi,\psi)$ can also arise for some values of $A$ and $B$. In this regard, one can refer to the observations, since mean parameter values of the hybrid expansion law from the $H(z)+\text{SN Ia}$ data are: $A\approx 0.488$ and $B\approx 0.444$ \cite{akarsu2014}. Note that for these parameters the discontinuities does not arise. 

After obtaining the scalar-tensor model for the two stages of cosmic history, we have focused on recovering geometric ({\it i.e.} standard) formulation of the $f(\mathcal{G},T)$ gravity. In order to achieve this, we have considered the de-Sitter (late time) evolution of the Universe since results have been obtained analytically. In what follows, for the potential (\ref{eq43}) where $w=0$, the recovered $f(\mathcal{G},T)$ function is found to be $f({\mathcal{G}},T)=-V_0+f_1(\mathcal{G})\pm \psi_0(-\rho_0 T)^{1/2}$ with the $f_1(\mathcal{G})$ being the arbitrary function of the GB term. Remarkably, the $f_2(T)$ part takes the same form as in the scalar-tensor $f(R,T)$ gravity \cite{gonccalves2022}. Hence, both theories are closely connected and can be distinguished only by the terms involving the scalar-curvature coupling. Our work is closed with the reconstruction of the quadratic potential $V(\phi,\psi)=\alpha \phi+\beta \phi^2 -\gamma \psi^2$, for which the corresponding geometric formulation has a function $f(\mathcal{G},T)=\frac{\alpha^2}{4\beta}+ \frac{1}{2\beta}(\frac{1}{2}\mathcal{G}^2-\alpha\mathcal{G})-\frac{T^2}{4\gamma}$. As a consequence, if the general conditions are met and the potential $V(\phi,\psi)$ as well as the fields $\phi$ and $\psi$ are invertible, one can get corresponding $f(\mathcal{G},T)$ formulation without complications. Moreover, since the change between formulations is independent explicitly on the scale factor $a(t)$, one can transform between formulations anytime. This is interesting from a practical manner, as some tasks can be easier in one of those formulations than in the other.

As a conclusion, we have successfully developed the scalar-tensor counterpart of the well-established $f(\mathcal{G},T)$ gravity, where GB term is responsible for the geometric part \cite{sharif2016,yousaf2020,maurya2020,zubair2021,sharif2022,sharif2022b}. The resulting formulation is characterized by the two scalar fields $\phi$ and $\psi$ coupled to the GB term and trace of the energy-momentum tensor, respectively. The scalar-tensor representation introduced here can be regarded as an extension of the scalar Einstein-Gauss-Bonnet (EGB) theory by integrating field coupled with matter. Note that in general additional fields in modified gravity theories are often challenging to interpret physically \cite{clifton2012,nojrii2017}. However, in the context of the $f(\mathcal{G},T)$ gravity interpretation becomes more natural. Specifically, $\phi$ can be regarded as a dilaton akin to that in the scalar-EGB gravity \cite{ripley2019,nojrii2017,lyu2022}. Moreover, $\psi$ may conceptually resemble a massless (particularly at large-structure scales) form of a chameleon field, given its direct coupling with a matter \cite{clifton2012,khoury2013,betz2022}. We also remark, that for the chameleon-$\psi$ association, a more natural choice would be the scalar-tensor counterpart of the $f(\mathcal{G},L_m)$ theory, since generally the chameleon field couples with the matter Lagrangian \cite{khoury2013}.

To this end, the present study delves into a specific cosmic scenario linked to the hybrid scale factor proposed in \cite{akarsu2014}, by describing two distinct epochs in the history of the Universe. Our findings suggest that within the framework introduced here, cosmological reconstruction methods successfully capture various cosmic evolutions if the $\phi$ and $\psi$ scalar fields are well-behaved and invertible. Thus, as a next research direction, it will be interesting to study other cosmic scenarios such as the inflation or bounce \cite{linde2007,nojrii2017}. In order to better grasp the validity of the theory, the scalar-tensor formulation described here should be confronted with a different GR tests, which every modified gravity should pass \cite{ishak2019}. The theory should be also constrained from the energy conditions perspective \cite{sharif2016} as they can help with ruling out the physically unrealistic models. Additionally, given its inherent link between geometry and matter, the study of the particle creation and baryogenesis epoch in our framework may be really beneficial. The same stands for the comparison between the scalar-tensor $f(\mathcal{G},T)$ and the $f(R,T)$ gravities \cite{davoudiasl2004,lobo2022,pinto2022}. Furthermore, to gain more comprehensive understanding of the role played by the scalar-tensor $f(\mathcal{G},T)$ gravity in cosmology, our upcoming investigations will be focused on comparing this theory with a current observational data. Specifically, our next phase of research aims to constrain this theory by using various datasets, such as the $Pantheon+$ or $BAO$ ones \cite{koyama2016}. Therefore, crucial cosmological and astrophysical tests are yet to be conducted.

\bibliographystyle{apsrev}
\bibliography{biblo}

\end{document}